\documentclass{jfm}

\usepackage{graphicx,color}
\usepackage{float}
\usepackage{epstopdf, epsfig}
\usepackage{caption,subcaption}
\usepackage[intlimits]{amsmath}
\usepackage{lscape}

\newcommand\Web{\mbox{\textit{We}}}
\newcommand\Mar{\mbox{\textit{Ma}}}

\shorttitle{Impact of droplets onto surfactant-laden thin liquid films}
\shortauthor{C. R. Constante-Amores  et al.}

\title{
Impact of droplets onto  surfactant-laden thin liquid   films
}

\author{C. R. Constante-Amores$^{1,2}$\corresp{\email{crc15@ic.ac.uk}},
L.Kahouadji$^2$, S. Shin$^3$, J. Chergui$^4$, D. Juric$^{4,5}$,
\ns J. R. Castrej\'on-Pita$^{6}$,
\ns O. K. Matar$^{2}$,
\ns A. A. Castrej\'on-Pita$^{1}$
}
  
\affiliation{
$^1$Department of Engineering Science, University of Oxford, Oxford OX1 3PJ, UK\\[\affilskip]
$^2$ Department of Chemical Engineering, Imperial College London, London SW7 2AZ, UK
\\[\affilskip]
$^3$Department of Mechanical and System Design Engineering, Hongik University, KR
\\[\affilskip]
$^4$ Universit\'e Paris Saclay, Centre National de la Recherche Scientifique (CNRS), Laboratoire Interdisciplinaire des Sciences du Num\'erique (LISN), 91400 Orsay, France
\\[\affilskip]
$^5$Department of Applied Mathematics and Theoretical Physics, University of Cambridge, Cambridge CB3 0WA, UK
\\[\affilskip]
$^6$Department of Mechanical Engineering, University College London, WC1E 7JE, United Kingdom
}

\begin{document}

\maketitle

\begin{abstract}
\noindent
We study the effect of insoluble surfactants on the impact of surfactant-free droplets on surfactant-laden thin liquid films via a fully three-dimensional direct numerical simulations approach that employs a hybrid interface-tracking/level-set method, and by taking into account surfactant-induced Marangoni stresses due to gradients in interfacial surfactant concentration. Our numerical predictions for the temporal evolution of the surfactant-free crown are validated against the experimental work by \citet{che_matar}. We focus on the `crown-splash regime', and we observe that the crown dynamics evolves through  various stages: from the  the growth of linear modes (through a Rayleigh-Plateau instability) to the development of nonlinearities leading to primary and secondary breakup events (through droplet shedding modulated by an end-pinching mechanism). We show that the addition of surfactants does not affect the wave selection via the Rayleigh-Plateau instability. However, the presence of surfactants plays a key role in the late stages of the dynamics as soon as the ligaments are driven out from the rim. Surfactant-induced Marangoni stresses delay the end-pinching mechanisms to result in longer ligaments prior to their capillary singularity.  Our results indicate that Marangoni stresses bridge the gap between adjacent protrusions promoting its collision and the merging of ligaments. Finally, we demonstrate that the addition of  surfactants leads to surface rigidification and consequently to the retardation of the flow dynamics. 
\end{abstract}

\section{Introduction}
Droplets impacting on liquid surfaces are observed in a broad range of natural phenomena and industrial applications. Consequently, its study is driving a significant scientific interest; for comprehensive reviews please see e.g., \citet{yarin,Josserand,Cheng}. The complex topological features of droplet impact have attracted the fluid mechanics community since the first  ground-breaking experiments from \citet{Worthington} and \citet{Edgerton} over a century ago.  However, it was not until the advent of high-speed imaging and high-resolution computational tools that it became possible to capture the early dynamics of the broad range of scales involved in such phenomena \citep{thoroddsen_2002, Thoraval_prl,zhang_2012,agbaglah_deegan_2014,josserand_2016,Thoroddsen_2012,Fudge}.

The dynamics of droplet impact on liquids is extremely diverse and is determined by a wide number of factors including the impact speed and the fluid properties, as well as whether the impact occurs on either a deep pool or a thin layer of fluid \citep{josserand_2016,Deegan_2007,che_matar}. At high impact speeds, the latter situation results in the formation of an upward Worthington jet  \citep{gekle_gordillo_2010}, whereas the former case gives rise to the well-known  {\it Milkdrop Coronet}, which is the subject of this paper \citep{Deegan_2007,josserand_2016,thoroddsen_2002}. Instants after impact on to a thin liquid layer, inertia prompts the formation of an ejecta sheet. After some time, its edge forms a capillary-driven cylindrical-shaped rim that evolves through various stages: first, a combination of Rayleigh?Taylor and Rayleigh-Plateau instabilities growth on the rim (see  \citet{Agbaglah_2013,Zhang_RP}), to then develop non-linearities leading to breakup (through end-pinching, see \citet{wang_bourouiba_2018}), and the formation of a cascade of droplet sizes. 

Under most natural or industrial conditions, streams are contaminated with surfactants (i.e., surface-active agents), whose concentration variations lead to surface tension gradients,  which in turn, result in the formation of Marangoni stresses and surface viscosities (see \citet{manikantan_squires_2020}). Previous studies have demonstrated the role of surfactants in capillary singularities (see for instance, \citet{Ananthakrishnan}, \citet{craster_pof_2002}, \citet{Liao_2004},  \citet{Kamat_prf_2018} and \citet{Constante-Amores_prf_2020}); here, we study the role of surfactant-induced flows on the late stages of splashing. \citet{che_matar} demonstrated the role of surfactants on the impact dynamics onto thin liquid films for low and moderate Weber and high Reynolds numbers, e.g., $2.6<We<532$ and $3000<Re<8654$, respectively ($Re=\rho_lUD/\mu_l$ and ${We=\rho_l U^2 D/\sigma}$, where $D$, $\rho_l$, $\mu_l$ $\sigma$, and  $U$ stand for the droplet diameter, and its characteristic density, viscosity, surface tension, and velocity, respectively). In this past work, the authors showed that the presence of surfactants affects the post-impact dynamics such as the propagation of capillary waves, the growth of the crown, and the generation of secondary droplets. However, Marangoni stresses, surfactant concentration dynamics and other important insights of the impact phenomena were not investigated. 

The vast majority of past works have focused on the physical understanding of the early dynamics of the ejected sheet using axisymmetric simulations, which is a valid assumption in the limiting case of 
$t\ll D/U$ (where $t$ stands for time) -- see for example \citet{Josserand_2003, josserand_2016, agbaglah_deegan_2014}. In contrast, the study of the crown-splash regime requires the use of full three-dimensional (3D) numerical simulations in order to accommodate the natural occurrence of possible symmetry-breaking events, such as, the growth of transversal instabilities in the rim, or rim breakup into droplets. However, 3D simulations still remain a numerical challenge \citep{Gueyffier} and have not often been reported in the literature, \citep{Liang,reijers2019oblique,Chen_2020,Xavier}. 

The present study aims to unravel the role of surfactant-induced Marangoni stresses on the interfacial dynamics during the impact of a surfactant-free droplet on a surfactant-laden thin layer of the same liquid in the crown-splash regime. Here,  we perform 3D   calculations of splashing in the presence of surfactants by using a hybrid front-tracking method to track the interface location.  Section \ref{Numerical} presents the governing dimensionless parameters, the problem geometry, initial and boundary conditions, and the numerical validation. Section \ref{sec:Results} provides a discussion of the results and concluding remarks are given in Section \ref{sec:Con}.

%----------------------------------------------
\section{Problem formulation and numerical method\label{Numerical}}
%----------------------------------------------
%

High resolution simulations were performed by solving the two-phase incompressible Navier-Stokes equations with surface tension in a three-dimensional Cartesian domain $\mathbf{x} = \left(x, y, z \right)$ (see figure \ref{configuration}a).  The interface between the gas and liquid is described by a hybrid front-tracking/level-set method, where (insoluble) surfactant transport is resolved at the interface \citep{Shin_jcp_2018}. Here, and in what follows, all variables are made dimensionless (represented by tildes) using

\begin{equation}
\quad \tilde{\mathbf{x}}=\frac{\mathbf{x}}{D_o},
\quad \tilde{t}=\frac{t}{t_{r}}, 
\quad \tilde{\textbf{u}}=\frac{\textbf{u}} {U},
\quad \tilde{p}=\frac{p}{\rho_{\textcolor{black}{l}} U^2}, 
\quad \tilde{\sigma}=\frac{\sigma}{\sigma_s},
\quad \tilde{\Gamma}=\frac{\Gamma}{\Gamma_\infty},
\end{equation}
	
\noindent	
where, $t$, $\textbf{u}$, and $p$ stand for time, velocity, and pressure, respectively. The physical parameters correspond to the liquid density $\rho_l$, liquid viscosity, $\mu_l$, surface tension, $\sigma$, surfactant-free surface tension, $\sigma_s$, and $U$ is the drop impact velocity, and $D_o$ its initial diameter; hence, the characteristic time scale is $t_{r}= D_o/U$. The interfacial surfactant concentration, $\Gamma$, is scaled by the saturation interfacial concentration, $\Gamma_{\infty}$.  As a result of this scaling, the dimensionless equations read 
\begin{equation}\label{div}
 \nabla \cdot \tilde{\textbf{u}}=0,
\end{equation}
\begin{equation*}
\tilde{\rho} (\frac{\partial \tilde{\textbf{u}}}{\partial \tilde{t}}+\tilde{\textbf{u}} \cdot\nabla \tilde{\textbf{u}}) + \nabla \tilde{p}  =  -\frac{\textbf{i}_z }{Fr^2}~ + \frac{1}{Re}~ \nabla\cdot  \left [ \tilde{\mu} (\nabla \tilde{\textbf{u}} +\nabla \tilde{\textbf{u}}^T) \right ] +
\end{equation*}
\begin{equation}\label{NS_Eq}
+\frac{1}{\Web}~
\int_{\tilde{A}\tilde{(t)}} 
(\tilde{\sigma} \tilde{\kappa} \textbf{n} +   \nabla_s  \tilde{\sigma})  \delta \left(\tilde{\textbf{x}} - \tilde{\textbf{x}}_{_f}  \right)\mbox{d}\tilde{A},
\end{equation}
 \begin{equation} 
 \frac{\partial \tilde{\Gamma}}{\partial \tilde{t}}+\nabla_s \cdot (\tilde{\Gamma}\tilde{\textbf{u}}_{\text{t}})=\frac{1}{Pe_s} \nabla^2_s \tilde{\Gamma}, 
 \end{equation}
\noindent
where the density and viscosity are given by $\tilde{\rho}=\rho_g/\rho_{\textcolor{black}{l}} + \left(1 -\rho_g/\rho_{\textcolor{black}{l}}\right) \mathcal{H}\left(\tilde{\textbf{x}},\tilde{t}\right)$ and $\tilde{\mu}=\mu_g/\mu_{\textcolor{black}{l}}+ \left(1 -\mu_g/\mu_{\textcolor{black}{l}}\right) \mathcal{H}\left( \tilde{\textbf{x}},\tilde{t}\right)$ wherein $\mathcal{H}\left( \tilde{\textbf{x}},\tilde{t}\right)$ represents a Heaviside function, which is zero in the gas phase and unity in the liquid phase, while the subscript `$g$' \textcolor{black}{`$l$'} designate the gas, \textcolor{black}{and liquid phases, respectively,} and $\tilde{\textbf{u}}_{\text{t}}= \left ( \tilde{\textbf{u}}_{\text{s}} \cdot \textbf{t} \right ) \textbf{t}$ is the tangential velocity at the interface in which $\tilde{\textbf{u}}_{\text{s}}$ represents the interfacial velocity, and $\kappa$ is the curvature. The interfacial gradient is given by $\nabla_s=\left({\mathbf{I}}-\mathbf{n}\mathbf{n}\right)\cdot \nabla$ wherein $\mathbf{I}$ is the identity tensor and $\mathbf{n}$ is the outward-pointing unit normal. In addition, $\delta$ is a Dirac delta function, equal to unity at the interface and zero otherwise, and $\tilde{A} (\tilde{t})$ is the time-dependent interface area. 
As justified by the final paragraph of \citet{Stone}, in our frame of reference $\textbf{u} \cdot \textbf{n}=0$, which gives rise to equation 2.4. The dimensionless groups that appear in the governing equations are defined as
\begin{equation}
Re =\frac{\rho_l U D_o}{\mu_l}, ~~~
 We =\frac{\rho_l U^2 D_o}{\sigma_s},  ~~~
Fr =\frac{ U }{\sqrt{g D_o}},  ~~~
Pe_s=\frac{ U D_o}{\mathcal{D}_s},~~~
\beta_s= \frac{\Re \mathcal{T} \Gamma_\infty}{\sigma_s}, 
\end{equation}
where $Re$, $We$, $Fr$, and $Pe_s$ stand for the Reynolds, Weber, Froude, and (interfacial) Peclet numbers.   Gravity is negligible during the impact as indicated by the Froude number value, i.e. $Fr \sim \mathcal{O}(10^2)$ (similar assumptions were made by \citet{Deegan_2007}).
The parameter $\beta_s$ is the surfactant elasticity number that is a measure of the sensitivity of the surface tension, $\sigma$, to the interface surfactant concentration, $\Gamma$. Here, $\Re$ is the ideal gas constant value ($\Re = 8.314$ J K$^{-1}$ mol$^{-1}$),  $\mathcal{T}$ denotes temperature,  and $\mathcal{D}_s$ stands for the interfacial diffusion coefficient.

The non-linear Langmuir equation is used to describe $\sigma$ in terms of $\Gamma$, this is  
\begin{equation}
    \tilde{\sigma}=1 + \beta_s \ln{ (1 -\tilde{\Gamma})}.
\end{equation}

\noindent
The Marangoni stress, $\tilde{\tau}$, is expressed as a function of $\tilde{\Gamma}$ as
\begin{equation}
\frac{1}{\Web} \nabla_s \tilde{\sigma}\cdot {\mathbf{t}}  \equiv
\frac{\tilde{\tau}}{\Web} = -
\frac{\Mar }{(1-\tilde{\Gamma})}  \nabla_s\tilde{\Gamma}\cdot {\mathbf{t}},
\end{equation}
\noindent
where $Ma=\beta_s/We=Re \mathcal{T} \Gamma_\infty/\rho_l U^2 D_o$ is the Marangoni parameter and $\mathbf{t}$ is the unit tangent to the interface. Finally, a dimensionless thickness of the liquid layer, $h=H/D_o$ is defined as the ratio between the liquid film thickness and the droplet diameter. Tildes are dropped henceforth. 

%----------------------------------------------
\subsection{Numerical Setup, validation and parameters}
%----------------------------------------------

\begin{figure}
\begin{center} 
\includegraphics[width=1.0\linewidth]{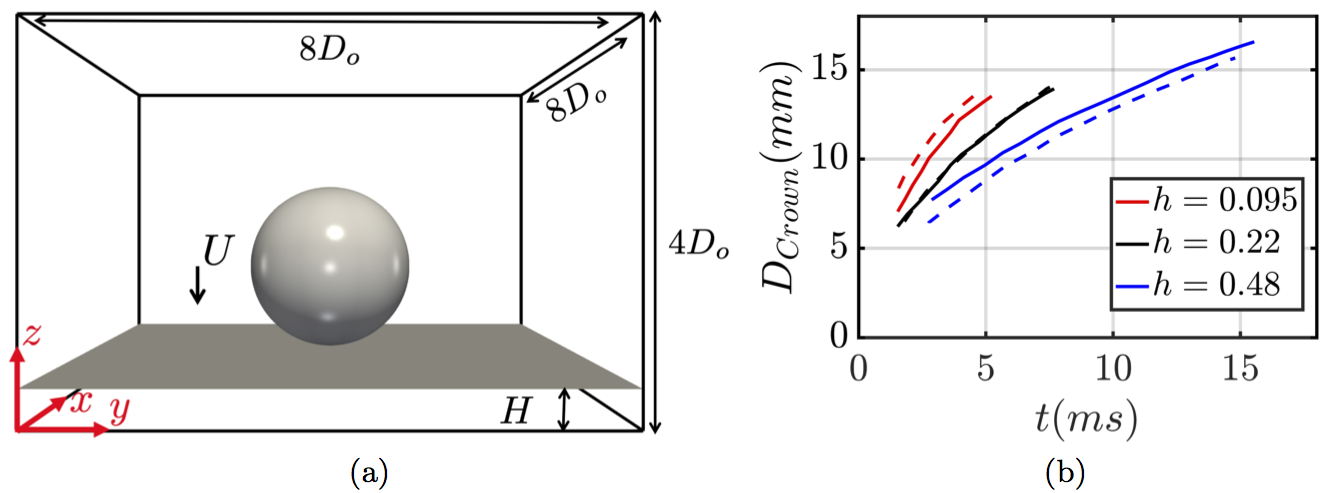} 
\end{center} 
\caption{\label{configuration} Schematic representation of the flow configuration, and validation of the numerical procedure: (a) initial configuration, highlighting the computational domain of size  (not-to-scale) $(8D_o \times 8D_o \times 4D_o)$ in a three-dimensional Cartesian domain, $\mathbf{x} = (x, y, z)$, with a resolution of $768^2\times 384$; (b) validation of the numerical results (solid lines) against the experimental data (dashed lines) for the crown rim diameter temporal evolution of a surfactant-free case reported in \citet{che_matar}: effect of varying the dimensionless film depth, $h=H/D_o$, while keeping the other parameters constant ($Re=7514$, $We = 249$, and $Fr = 13.65$).} 
\end{figure}

Figure \ref{configuration}a highlights the geometry and the initial and boundary conditions of the problem, which closely follows previous work by \citet{josserand_2016,agbaglah_deegan_2014}, and \citet{Fudge}. A drop of initial size $D_o$ with velocity $U$ impacts a uniform layer of thickness $H$ of the same liquid. The size of the computational domain is  $8D_0 \times 8D_0 \times 4D_0$, which is sufficiently large to avoid any effects of artificial reflections from the boundaries. The centre of the drop is located at a small distance above the pool surface (e.g., an initial separation of $0.05D_0$). A no-slip boundary \textcolor{black}{and no-penetration conditions are}  assumed for the bottom wall of the domain, whereas a no-penetration boundary condition is prescribed for the top and lateral domain boundaries (similar to previous work by \citet{batchvarov2020threedimensional}).

Figure \ref{configuration}b illustrates the ability of the numerical technique to reproduce the temporal evolution of the crown experimentally obtained and reported by \citet{che_matar}. After impact, inertia drives the  rapid ejection of a vertical sheet from the pool, then capillarity prompts the formation of a rim, and then a crown.  We observe excellent agreement for $h=0.22$, while a small offset is observed for other two film thicknesses. The discrepancy could be attributed to conditions found in experimental conditions but not considered here, such as the surface waves produced during drop formation or air-induced flows during droplet fall. Nonetheless, the agreement between the numerical method and experimental results is visible. Additional validation cases are found in Appendix A. The numerical technique has also been   validated for drop-interface coalescence, which can be considered as a limiting case of drop impact onto a pool, i.e., $U=0$ (see \citet{constante_coales} for more details). The nonlinear interfacial dynamics have been validated for the capillary breakup of liquid threads  (see \citet{Constante-Amores_prf_2020,constante_jets} for more details). The numerical validation of the surfactant equations has been previously presented in \citet{Shin_jcp_2018}. 

In terms of mesh resolution, we have ensured that our numerical simulations are mesh independent, and as a consequence, the numerical findings do not change by decreasing the cell size for a resolution of $768^2 \times 384$ (i.e.,  \textcolor{black}{$D_0/\Delta {\bf x}=96$} cells, see Appendix B). Additionally,  liquid volume and surfactant mass conservation are met under errors of less than $10^{-1}\%$, and $10^{-2}\%$, respectively.  Extensive mesh studies for surface-tension-driven phenomena using the same computational method can be found elsewhere, e.g. \citep{batchvarov2020threedimensional,batchvarov2020effect,Constante-Amores_prf_2020,constante_jets,sheet_constante,constante_2023}.

The dimensionless values for the investigated phenomenon are consistent with experimentally realisable systems. We assume a water-air system, where the density and viscosity ratios are $\rho_g /  \rho_l = 1.2 \times 10^{-3}$ and $\mu_g / \mu_l = 0.018$, respectively. The targeted flow conditions for the surfactant-free base case are based  on \citet{Deegan_2007} who built a phase diagram in the $We-Re$ space for $h=0.2$. Our study focuses on  the `crown-splash' regime (i.e., $We>500$ and $Re <1100$), as we aim to stay  away from the `microdroplet-splash' regime (i.e., $We>500$ and $Re <1000$). The latter regime is characterised by the formation  of a rapid ejecta sheet which  undergoes capillary singularities to  form  microdroplets, requiring  an extremely large resolution to capture  droplet formation  from  the ejecta sheet and from the crown. Thus, we have selected $Re=1000$, $We=800$ and $h=0.2$ as the targeted surfactant-free case. 

We have examined the case of adding surfactant to the thin liquid film but not the impacting droplet as it is less challenging   to control  the characteristics of  surfactant-free drops experimentally. We have considered insoluble surfactants whose critical micelle concentration (CMC), i.e. $\Gamma_\infty \sim \mathcal{O}(10^{-6})$ mol m$^{-2}$  for NBD-PC (1-palmitoyl-2-{12-[(7-nitro-2-1,3-benzoxadiazol-4-yl)amino]dodecanoyl}-sn-glycero-3 -phosphocholine); thus, we have explored the range of $0.1 <\beta_s<0.9$ which corresponds to $ \mathcal{O}(10^{-7})<$CMC$<\mathcal{O}(10^{-6})$ mol m$^{-2}$, for typical values of $\sigma_s$. We have set $Pe_s=10^2$ following  \citet{batchvarov2020effect} and \citet{Constante-Amores_prf_2020} who \textcolor{black}{demonstrated} that the interfacial dynamics are weakly-dependent on $Pe_s$ beyond this value.

For a water-droplet of a typical size of $D_o \sim \mathcal{O}(10^{-3})$m, the impact time scale $T_{imp}=t_r=D_o/U \sim \mathcal{O}(10^{-4}-10^{-3})$s; whereas the Marangoni time-scale  $T_{\tau}$ can be estimated by a balance between the Marangoni  and the viscous stresses, resulting in  $T_{\tau} \sim \mu_l D^2 / (h \Delta \sigma) \sim \mathcal{O}(10^{-4}-10^{-3})$s (see \citet{che_matar} for more details).  Therefore, surfactant-induced Marangoni stresses will play a crucial role in the flow physics. A discussion of the results is presented next.

%----------------------------------------------
\section{Results\label{sec:Results}}
%----------------------------------------------

\begin{figure}
\begin{center} 
\includegraphics[width=1.0\linewidth]{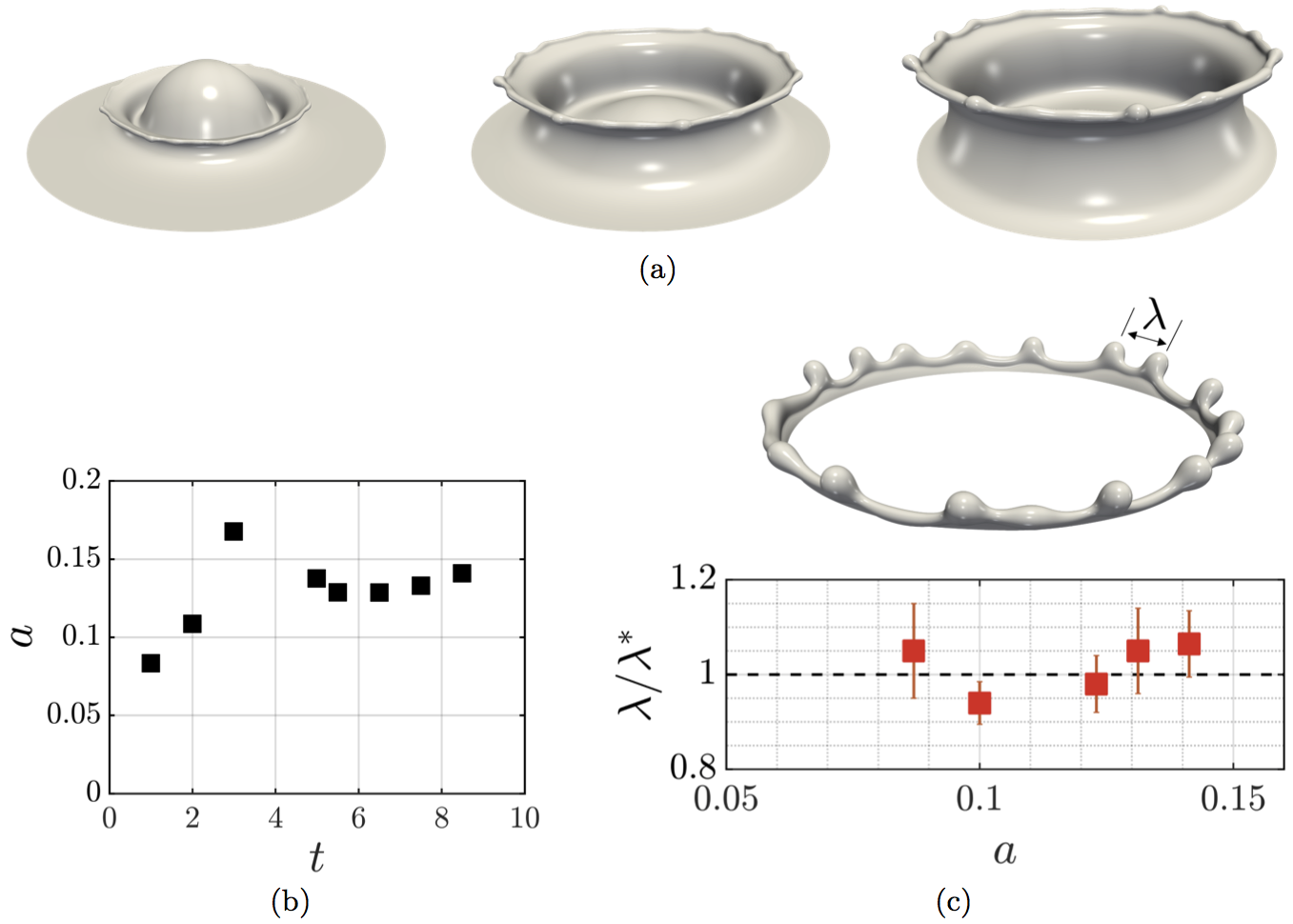} 
\end{center} 
\caption{
Wavelength selection in the crown splash regime. Panel (a) presents the  early interfacial  dynamics  through a  three-dimensional representation of the predicted interface location for $Re=1000$ and $We=800$, at $t=(0.5, 1.0, 1.5)$ corresponding to columns one to three, respectively. Panel (b) plots the temporal evolution of the rim radius. Panel (c) shows the selection of the wavelength    $\lambda$ normalized with the theoretical RP-instability, $\lambda^*$ as a function of the local rim radius, $a$, at early times times of the simulation for $t<5$. A three-dimensional representation of the crown for  the dimensionless time $t=7.5$ is also presented, highlighting the predicted wavelength between crests. 
 \label{rim}} 
\end{figure}

First, we discuss the early-time dynamics of the surfactant-free impact at $Re=1000$ and $We=800$.  Figure \ref{rim}a shows the three-dimensional representation of the interface at early times, i.e. prior transversal instabilities grow at the rim. As seen in this figure, instants after impact, we observe the formation of an axisymmetric thin layer of ejecta fluid which draws fluid from the droplet eventually giving rise to a {\it Peregrine sheet}, in agreement with \citet{Deegan_2007} and \citet{josserand_2016}. As time evolves, the  Peregrine sheet grows upwards and  expands radially while its film thickness reduces; here the radial coordinate $r$ is defined at the initial point of impact. A closer inspection of the interfacial shape of the ejecta sheet  at $t=0.5$ \textcolor{black}{demonstrates} that its orientation is nearly horizontal, i.e. parallel to the pool. We also see surface tension forces inducing the formation of a hemispherical tip at the edge of the ejecta, which leads to a pressure gradient between the tip and the adjacent sheet, which results in fluid motion from the tip towards the sheet.  This behaviour leads to the formation of a sheet perpendicular to the pool at longer times; this in agreement with the experimental observations by \citet{Zhang_RP} and \citet{Deegan_2007}.
Figure \ref{rim}b below shows the  temporal evolution of the rim size  for the surfactant-free case at the crossing of the $x-z$ plane ($y=4.90$). The peak observed between $t=3-5$ is due to the growth of corrugations in the rim. As the rim elongates to form the crown, we observe the growth of an azimuthal instability on the surface. Our results indicate that these naturally-occurring rim instabilities select wavelengths that are consistent with the most unstable Rayleigh-Plateau (RP) instability.
We have measured the average distance ($\lambda$) between the `undulations' or crests seen along the rim. We then normalised the result by the most unstable Rayleigh-Plateau (RP) instability wavelength, given by $\lambda^*= 2 \pi /k= 9.016 a$, where $k$ and $a$ are the wavenumber of the fastest-growing instability mode, and the rim radius, respectively \citep{Agbaglah_2013,Zhang_RP}. In Figure \ref{rim}c we plot the selected wavelength as a function of the rim radius; for the cases studied here, $\lambda/\lambda^* = 1.0 \pm 0.1$. These results are in agreement with the experimental results from \citet{Zhang_RP} and the theoretical and computational work from \citet{Agbaglah_2013} and \citet{sheet_constante}. A closer inspection of the top panel of figure  \ref{rim}c  shows that the interfacial corrugations are not always evenly distributed along the rim, as observed in experiments \cite{Thoraval_prl}. In particular, we observe large wavelengths, which are the result of interactions between adjacent instability modes, which is consistent with experimental observations. 

%%%%%%%%%%%%%%%%%
%%%%%%%%%%%%%%%%%
%%%%%%%%%%%%%%%%%
%%%%%%%%%%%%%%%%%
%%%%%%%%%%%%%%%%%
%%%%%%%%%%%%%%%%%

\begin{figure}
\begin{center} 
\includegraphics[width=1.0\linewidth]{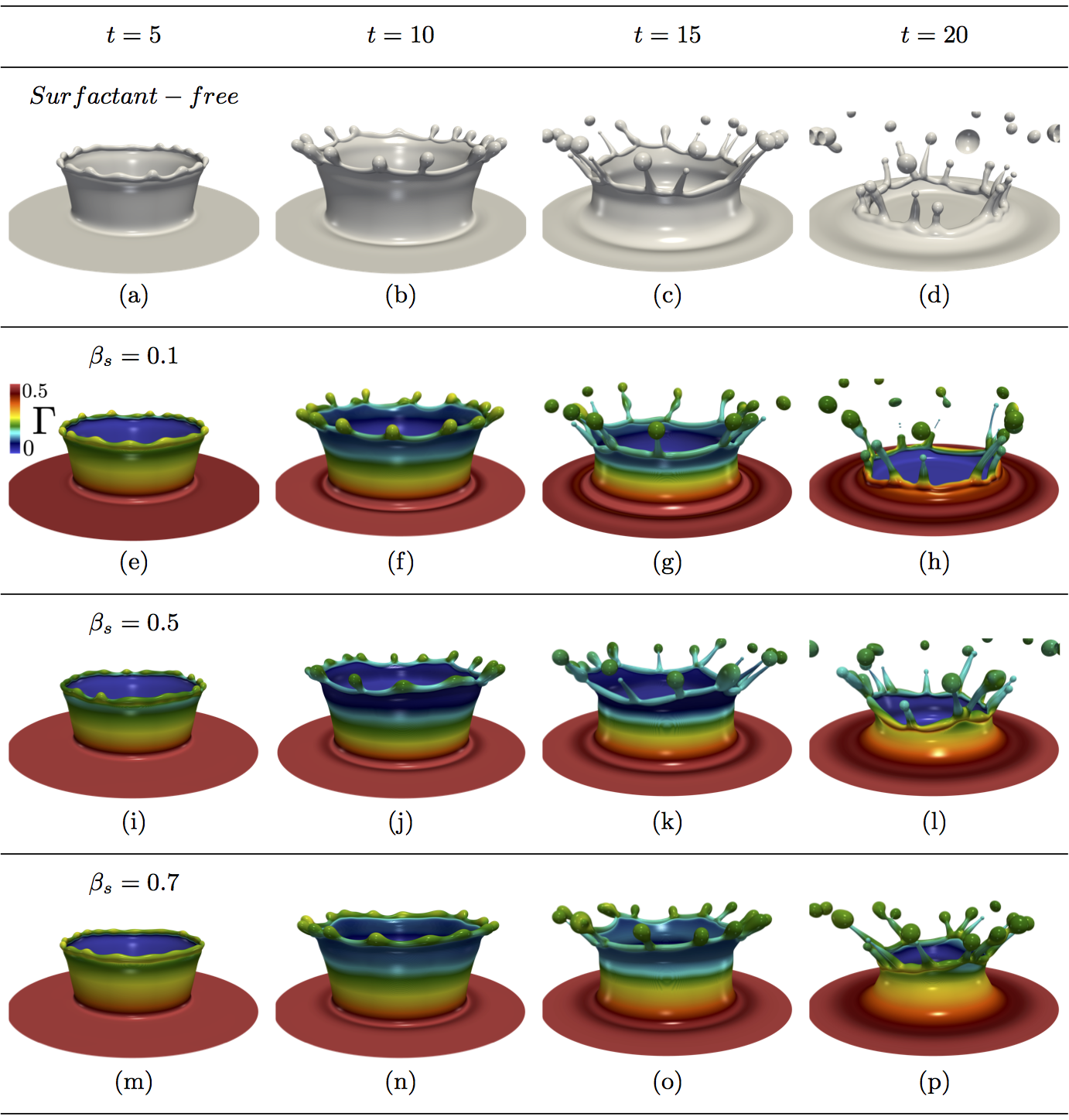} 
\end{center} 
\caption{\label{temporal_evolution} 
Effect of $\beta_s$ on the drop impact  dynamics for insoluble surfactants. Spatio-temporal evolution of the three-dimensional interface shape for surfactant-free cases, (a)-(d), and surfactant-laden cases for $\beta_s = (0.1, 0.5, 0.7)$ corresponding to panels (e)-(h), (i)-(l) and (m)-(p), respectively. Here, the dimensionless parameters are $Re=1000$, $We=800$ and $h=0.2$. For the surfactant-laden cases, $Pe_s=100$ and $\Gamma=\Gamma_\infty/2$, the colour indicates the value of $\Gamma$, and the legend is shown in panel (e).} 
\end{figure}

Figures \ref{temporal_evolution}(a-d) show a three-dimensional representation of the temporal evolution of the interface, from the time instabilities become visible to the time droplets breakup by end-pinching. At these longer times ($t>5$), the rim is weakly affected by the dynamics occurring at the impacting point, and interfacial RP instabilities trigger the development of ligaments (see figure \ref{temporal_evolution}a). With increasing time, liquid ligaments grow perpendicular to the rim resulting in a local  pressure-gradient-induced flow from the ligament to the tip, triggering the formation of  a  capillary-induced blob and a neck (displayed on figure \ref{temporal_evolution}b). The neck thins and stretches, due to a capillary-driven flow, to eventually resulting in the capillary pinch-off of a drop (the  well-known `end-pinching' mechanism proposed by \citet{stone_leal_1989}), see figure \ref{temporal_evolution}c. We note that the interfacial dynamics closely resembles experimental observations by \citet{Zhang_RP}, \citet{Josserand_2003}, and  \textcolor{black}{\citet{Deegan_2007}}, as well as the dynamics of retracting sheets reported by \citet{wang_bourouiba_2018} and \citet{sheet_constante}.

%%%%%%%%%%%%%%%%%
%%%%%%%%%%%%%%%%%
%%%%%%%%%%%%%%%%%
%%%%%%%%%%%%%%%%%
%%%%%%%%%%%%%%%%%
%%%%%%%%%%%%%%%%%

Next, we proceed to study the role of surfactants on the flow dynamics; as illustrated by figures \ref{temporal_evolution}(e-p) where we observe the droplet impact at various elasticity parameters $\beta_s$. As seen in these figures, regardless of the value of $\beta_s$, large surfactant concentration gradients are found on the outer interface of the ejecta sheet. This result supports the hypothesis that the ejecta sheet is generated at the interface of the surfactant-laden pool. On the other hand, the inner interface of the ejecta sheet arises from the surfactant-free drop, i.e., $\Gamma$ vanishes on the inner surface, see figure \ref{temporal_evolution}e. This observation, that the fluid source of the inner and outer surfaces of the ejecta sheet are the pool and the droplet, respectively, is in good agreement with the work by \citet{josserand_2016}. A close inspection of the distribution of $\Gamma$ in the outer sheet shows that $\Gamma$-values are maximum at the base of the ejecta, and their values reduce upwards (see for example panel (e) of figure \ref{temporal_evolution}). This is the result of an increase of the interfacial area, that leads to a dilution of surfactant at the interface, thus the surfactant concentration (surface tension) decreases (increases). In addition, the interfacial expansion results in large convective effects that transport surfactant towards the rim; see figure \ref{temporal_evolution}e, in which larger values of $\Gamma$ are found in the rim than in its adjacent sheet. 

The increase (decrease) of $\Gamma$ ($\sigma$) implies that a large surface deformation is required to satisfy the normal stress balance at the interface. Similar to the surfactant-free case, capillary-induced flow results in the formation of a rim on the sheet edge, whose thickening leads to the onset of destabilisation.  This is in agreement with previous studies from \citet{asaki}, who reported a surfactant-driven dampening effect on the capillary waves.

In addition, we expect higher (lower) values of $\Gamma$ at the rim wave crests (instability waves) as they belong to a radially converging (diverging) region, which have the effect of locally increasing (reducing) $\Gamma$. 
The gradients in $\Gamma$ result in surfactant-induced Marangoni stress flows from the {lower}-$\sigma$ radially-converging to the higher-$\sigma$ radially-diverging regions. This slows down the interfacial dynamics, and the formation of instabilities at the rim, in agreement with the idealised case presented by \citet{sheet_constante}. By increasing $\beta_s$ the surfactant distribution along the interface is enhanced, as displayed in figure \ref{temporal_evolution}i-l and figure \ref{temporal_evolution}j-m for $\beta_s=0.5$ and $\beta_s=0.7$, respectively.

\begin{figure}
\begin{center} 
\includegraphics[width=1.0\linewidth]{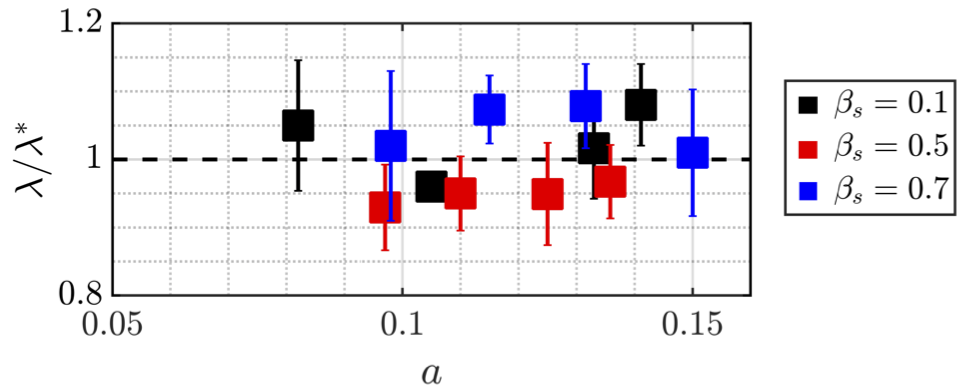} 
\end{center} 
\caption{\label{RP_surf} Effect of the elasticity parameter, $\beta_s$ in  the selection of  wavelength of the undulations normalized with the theoretical RP instability, $\lambda^*$, as a function of the local rim radius, $a$.} 
\end{figure}

Surfactant-induced Marangoni stresses only delay the development of the ligament from the rim; these eventually break up via end-pinching mechanism to form droplets, as shown in figure \ref{temporal_evolution}g, k,o. These results offer an explanation to the experimental observations of \citet{che_matar}. Figure \ref{RP_surf} \textcolor{black}{displays} the selected instability wavelength in terms of $\beta_s$ and $a_0$. Indeed, a close inspection indicates that the surfactant does not seem to greatly affect the selection of the wavelength; the predicted $\lambda$ is consistent with the most unstable RP instability at a ratio of  $\lambda/\lambda^* = 1.0 \pm 0.1$. Consequently, according to our results, the crown dynamics are mostly driven by the RP instability, although longer ligaments developed prior to break up in the presence of surfactants, as observed in figure \ref{temporal_evolution}\textcolor{black}{p}. This indicates that Marangoni stresses promote the \textcolor{black}{retardation} from end-pinching, \textcolor{black}{which, under certain circumstances may even lead to the neck re-opening, further delaying the splash,} a phenomenon previously demonstrated by \citet{kamat_2020, Constante-Amores_prf_2020, sheet_constante} 

\begin{figure}
\begin{center} 
\includegraphics[width=1.0\linewidth]{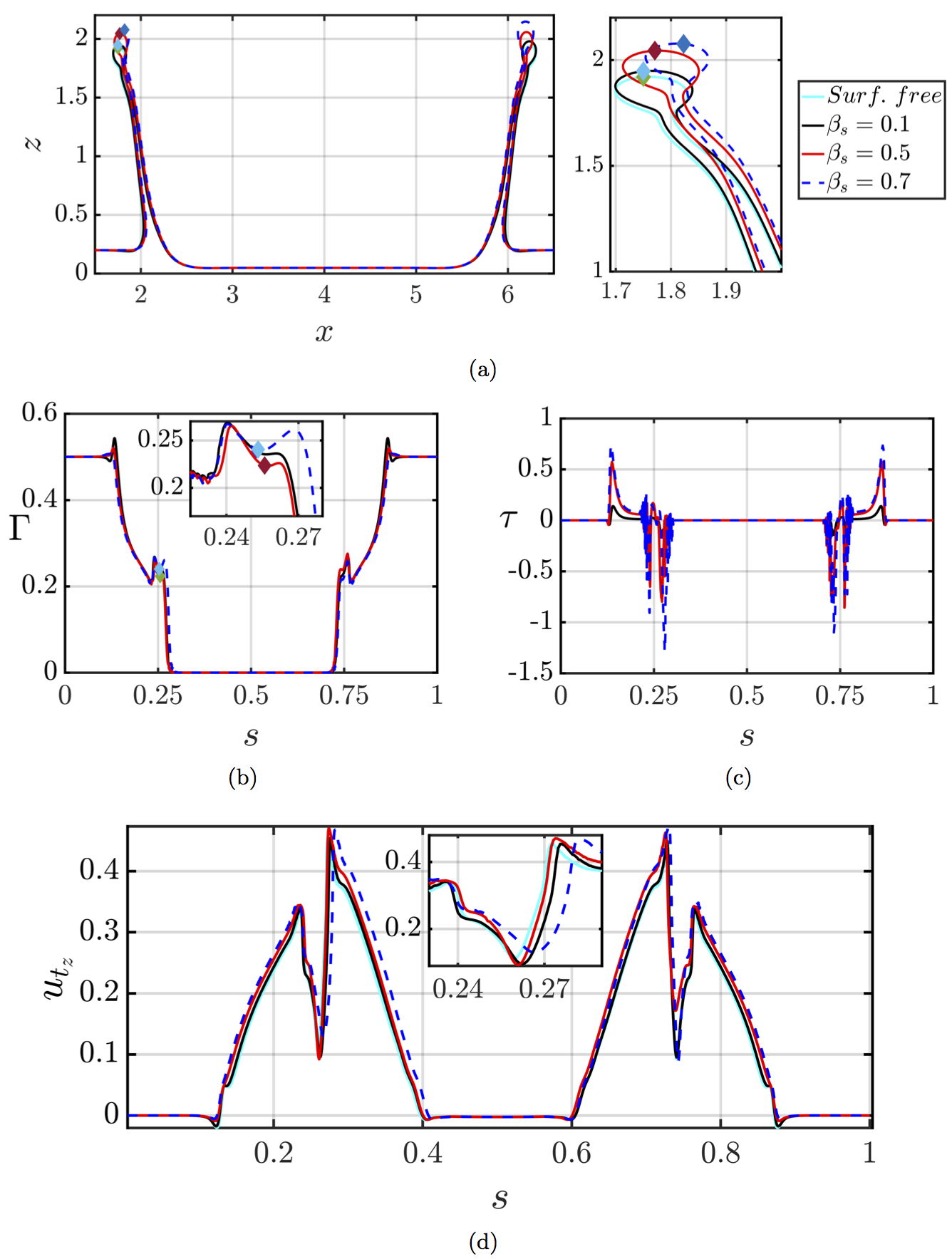} 
\end{center} 
\caption{\label{s_plots} Effect of the elasticity parameter, $\beta_s$, on the flow dynamics at $t=5$. Two-dimensional projections of the interface, $\Gamma$, $\tau$, and $u_{tz}$ in the $x-z$ plane ($y=4$) are shown in $(a)-(d)$, respectively.
In panel (a), a magnified view of the ejecta sheet is also presented. Note that the abscissa in $(a)$ corresponds to the $x$ coordinate, and in $(b)-(d)$ to the arc length, $s$. The arc length $s$ corresponds to the $x$?$z$ plane ($y = 4$) intersecting the interface, $s$ has been normalised on the full extent of $s$ associated with the length of the impact region in each case. The diamond shapes in panels $(a,b)$ indicate the location of the crown. All  parameters remain unchanged from figure \ref{temporal_evolution}.} 
\end{figure}

\begin{figure}
\begin{center} 
\includegraphics[width=1.0\linewidth]{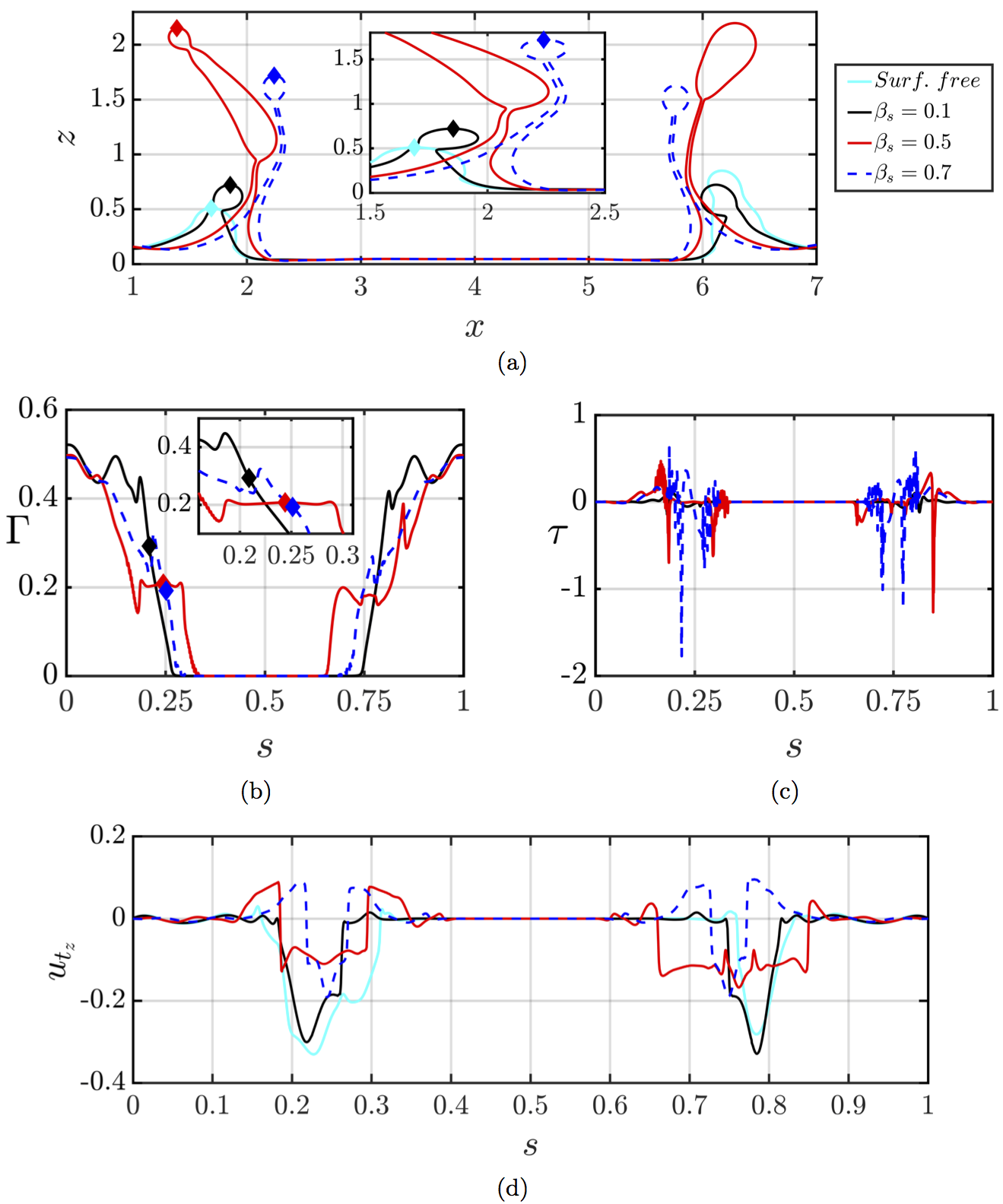} 
\end{center} 
\caption{\label{s_plots_t20} Effect of the elasticity parameter, $\beta_s$, on the flow dynamics at $t=20$. Two-dimensional projections of the interface, $\Gamma$, $\tau$, and $u_{tz}$ in the $x-z$ plane ($y=4$) are shown in $(a)-(d)$, respectively. Note that the abscissa in $(a)$ corresponds to the $x$ coordinate, and in $(b)-(d)$ to the arc length, $s$. The arc length $s$ corresponds to the $x$?$z$ plane ($y = 4$) intersecting the interface, $s$ has been normalised on the full extent of $s$ associated with the length of the impact region in each case. The diamond shapes in panels $(a,b)$ indicate the location of the crown. All  parameters remain unchanged from figure \ref{temporal_evolution}.} 
\end{figure}

Two-dimensional projections, in the $x-z$ plane ($y=4)$, of the interfacial shape, $\Gamma$, $\tau$, and the radial component of the interfacial velocity $u_{tz}$ in terms of $\beta_s$ and a fixed value of $Pe_s$, are displayed in figures \ref{s_plots} and \ref{s_plots_t20} corresponding to $t=5$ and $t=20$, respectively. As described above, at early times, i.e., $t=5$, the drop impact results in a non-uniform interfacial surfactant distribution, with high surface concentrations at the base of the ejected sheet. As seen, the interfaces of the surfactant-free and the $\beta_s=0.1$ surfactant-laden cases are practically indistinguishable. However, as the surfactant concentration increases, $\Gamma$ distributions trigger surfactant-driven dynamics from the sheet-base to its edge (see figure \ref{s_plots}b). As $\beta_s$ increases, long ejecta sheets are promoted resulting in the reduction of the rim and sheet thickness due to mass conservation (displayed in the magnified panel of figure \ref{s_plots}a). The increase in $\beta_s$ also enhances the $\Gamma$ distribution and increases the value of $\tau$ (see the increase of $\tau$ with $\beta_s$ in figure \ref{s_plots}c). Furthermore, Figure \ref{s_plots}c highlights the presence of large $\tau$  peaks at both sides of the rim due to the accumulation of $\Gamma$, and sudden drops at the base of the rim, suggesting a fluid transport from the rim to the sheet. In addition, a larger $\tau$ maximum is observed from the side of the inner sheet as the inner sheet is devoid of $\Gamma$ leading to an uneven flow motion inside the sheet. The variation of the streamwise interfacial tangential velocity $u_{t_{z}}$, along the arc length, is shown in figure \ref{s_plots}d. This figure, at the early stages of the dynamics, exhibits $u_t>0$ over the majority of the domain due to the dominance of the vertical radial expansion of the ejecta sheet.
A strong variation of $u_{t_{z}}$ is observed around the rim with  $u_{t_{z}}$ tending to zero at a sufficiently large distance away from the impact region. 

\begin{figure}
\begin{center} 
\includegraphics[width=0.7\linewidth]{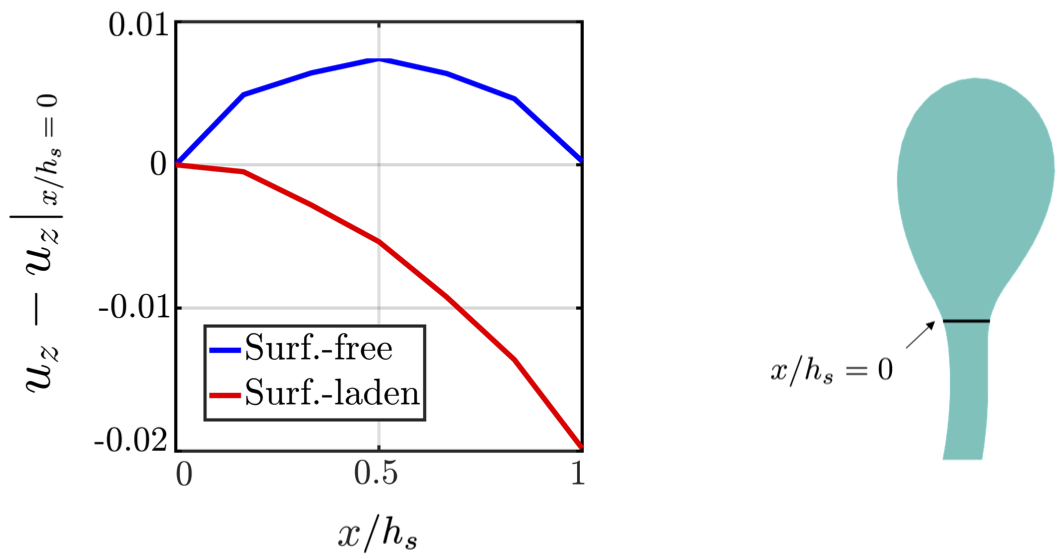} 
\end{center} 
\caption{\label{shear} Profiles of the velocity component, $u_z$, in the sheet-normal direction  inside  the sheet for the surfactant-free and surfactant-laden $(\beta_s=0.5)$ cases at $t=5$. 
Here, $u_z$ has been calculated in an inner-sheet frame-of-reference along the cross-stream direction, $x$, for the axial location along the rim neck. Note that the axial distance has been normalised with the average sheet thickness, i.e., $x/h_s$ (thus $x/h_s=0$ and $x/h_s=1$ correspond to the inner and outer sheet, respectively). All  parameters remain unchanged from figure \ref{temporal_evolution}.} 
\end{figure}

\begin{figure}
\begin{center} 
\includegraphics[width=1.0\linewidth]{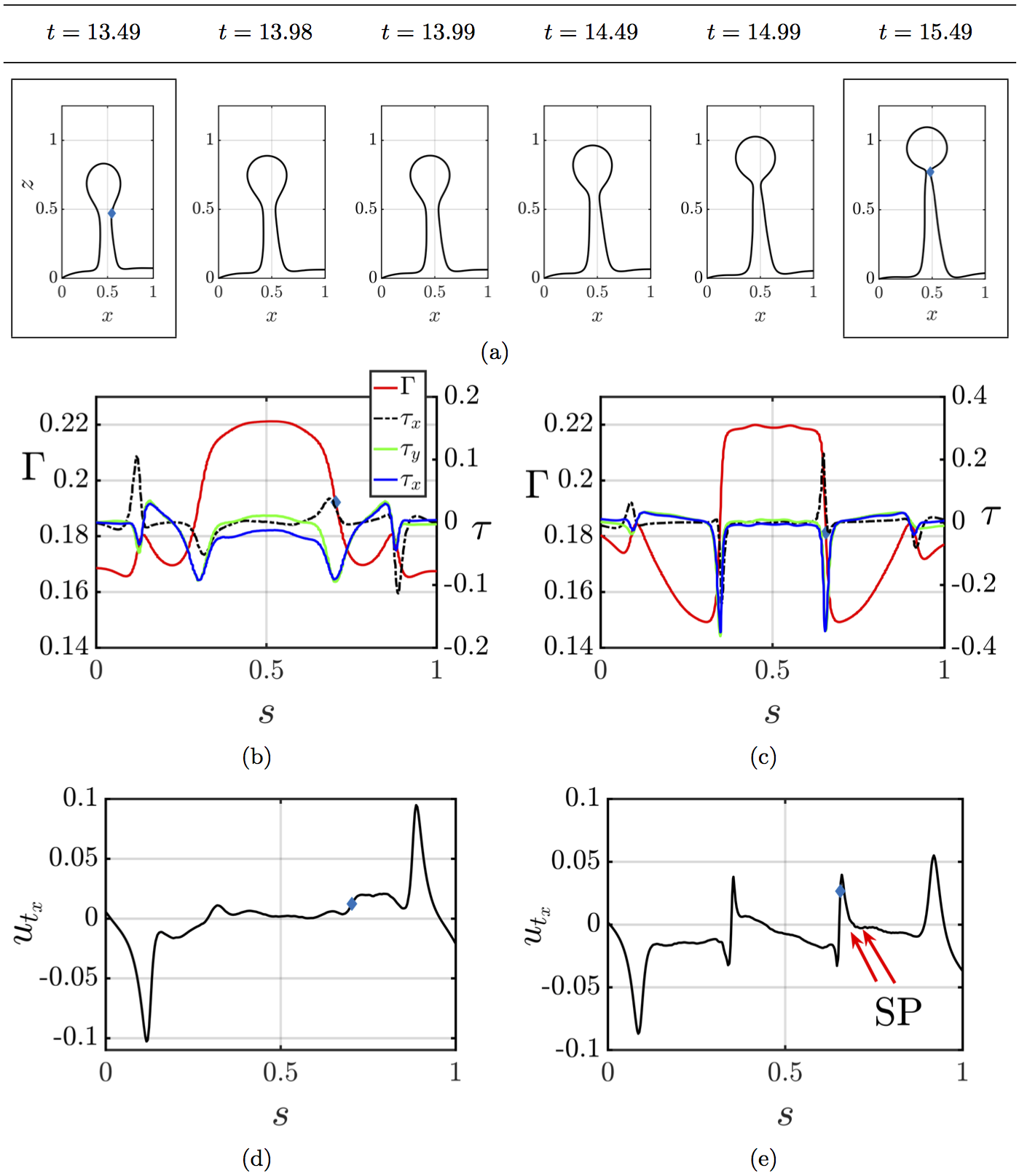} 
\end{center} 
\caption{\label{endpinching_figure} 
(a) Spatio-temporal evolution of a surfactant-laden ligament and its \textcolor{black}{retardation} from pinchoff driven by surfactant-induced Marangoni flow when $\beta_s=0.5$. (b) and (c) represent $\Gamma$ and $\tau$ as a function of the arc length $s$ for the framed panels of $(a)$, i.e.,  $t=13.49$ and $t=15.49$, respectively; (d) and (e) represent the tangential velocity $u_t$ as a function of the arc length $s$. The axial location  in $(a)$  has been normalised over the distance of the ligament on the axial direction; the \textcolor{black}{abscissa} direction has been normalised with respect to its value at $x=0$ for each \textcolor{black}{single} panel. The arc lengths have been normalised over  the full extent of $s$. The diamond markers  represent the location of the neck.  `SP' in (e) indicates the location of the stagnation points.  All  parameters remain unchanged from figure \ref{temporal_evolution}.} 
\end{figure}

The influence of $\beta_s$ at long times, $t=20$, is examined in  figure \ref{s_plots_t20}. Qualitatively, the surfactant effects are particularly evident as the dynamics have been slowed down by the surface rigidification brought by the Marangoni stresses, as can be seen in Figure \ref{s_plots_t20}d (the overall magnitude of $u_{t_z}$ is smaller with increasing $\beta_s$). As $\beta_s$ decreases ($\sigma$ increases), the surfactant is highly convected as $\tau$ tends to zero, lowering the interfacial tension and consequently increasing surface deformation. As  $\beta_s$  increases, the gradient in $\Gamma$ ($\tau$) decreases (increases) near the rim, and the Marangoni stresses predominantly retard the flow and ?rigidify? the interface.

As can be observed in the results presented in Figure \ref{s_plots_t20}c, an increase in $\beta_s$ results in the strengthening of $\tau$. Figure \ref{s_plots_t20}d provides conclusive evidence of the surface rigidification as most of the $u_{tz}$ is characterised by a negative value due to the decay of the dynamics. The variation of $u_{tz}$ decreases as $\beta_s$ increases, strengthening the  Marangoni stresses arising from the surfactant-induced surface tension gradients. Finally, according to the foregoing results, we conclude that the interfacial dynamics for surfactant-laden cases in the `crown-regime' resemble its surfactant-free counterpart except for the retardation due to the surface rigidification brought about by the presence of surfactant-induced Marangoni stresses.

In Figure \ref{shear}, we plot the component of the velocity variation in the sheet-normal direction, $u_z$, within the sheet, in a frame-of-reference moving with the inner-sheet at $t=5$. For the surfactant-free case, the liquid within the sheet follows a parabolic profile into the rim; this effectively corresponds to a no-slip condition with $u_z$ pinned to the interface, due to the absence of surfactants. In contrast, the retardation brought about by the presence of surfactants via Marangoni stresses results in shear flow for the surfactant-laden case. The higher surfactant concentration in the outer rim wall, results in   $\tau$ triggering the deceleration of the fluid entering into the bulbous near the outer wall. 

\begin{figure}
\begin{center} 
\includegraphics[width=0.6\linewidth]{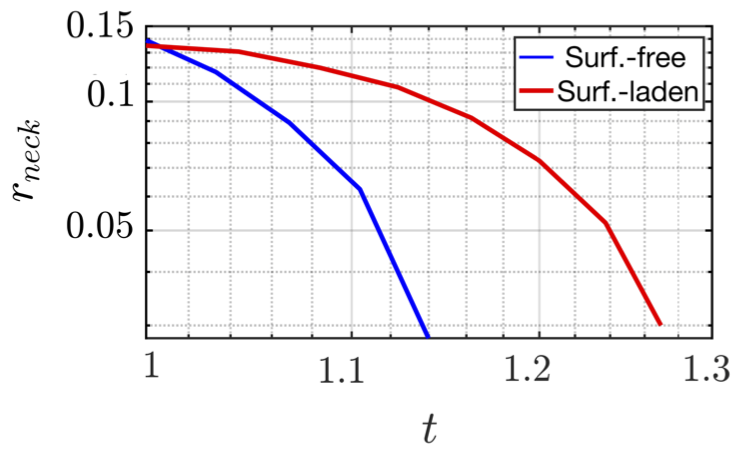} 
\end{center}
\caption{\label{neck_evolution} Temporal evolution of the neck radius for the surfactant-free and surfactant-laden case ($\beta_s=0.5$), highlighting the surfactant-driven retardation of the interfacial capillarity.}
\end{figure}

Figure \ref{endpinching_figure}a \textcolor{black}{presents} the late stages of a surfactant-laden ligament before the end-pinching mechanism. In the absence of surfactants, the competition between viscosity and capillarity drives the formation of and dynamics of bulbous edge on the ligament tip. We proceed to calculate the relative importance of viscous to surface tension forces via the local Ohnesorge number for the ligament (i.e. $Oh=\mu_l / \sqrt{\rho_l \sigma_s R}$, here $R$ stands for the  average radius of the ligament) to explain the retardation phenomenon observed in figure \ref{endpinching_figure}a. The resulting local  $Oh \sim 0.07$ (e.g., low $Oh$-regime), and subsequently the local flow dynamics affecting the ligament (i.e., surfactant-driven retardation from end-pinching) can be explained by the mechanism proposed by \citet{Constante-Amores_prf_2020}, \citet{kamat_2020} and \citet{Hoepffner_jfm_2013}. Figures \ref{endpinching_figure}b,c and \ref{endpinching_figure}d,e show $\Gamma$ and $\tau$, and the tangential velocity components $u_t$ for the framed panels in figure \ref{endpinching_figure}a, respectively. Figure \ref{endpinching_figure}b,d \textcolor{black}{demonstrate} that, near the neck (marked with a blue diamond symbol) $\tau$ results in $u_{tx}>0$ preventing the formation of a second stagnation point required to trigger the capillary singularity. As shown by \citet{Constante-Amores_prf_2020}, two stagnation points sandwiched between the neck boundaries are needed to induce end-pinching, as can be observed in \ref{endpinching_figure}d-e (e.g `SP' stands for stagnation points). Surface tangential stresses prevent capillary breakup for a longer time than in the surfactant-free case leading to longer ligaments and an increase in surface area. This results in the dilution (increase) of surfactant (surface tension) and the eventual reduction in the strength of surfactant-induced Marangoni stresses thus initiating end-pinching.

Figure \ref{neck_evolution} presents the temporal evolution of the neck radius for the surfactant-free and surfactant-laden cases (we cannot report the late stages of the neck due to the lack of snapshots up to its capillary singularity). As seen, at short times, the presence of surfactants at the interface leads to the reopening of the neck ($dr/dt \sim 0$), and subsequently, to a short period in which surfactants induce \textcolor{black}{retardation} from end-pinching. At a later time, as surfactants are evacuated from the neck, the dynamics eventually results in the interfacial singularity.

%%%%%%%%%%%%%%%%%%%%%%%%%%%%%
%%%%%%%%%%%%%%%%%%%%%%%%%%%%%

\begin{figure}
\begin{center} 
\includegraphics[width=1.0\linewidth]{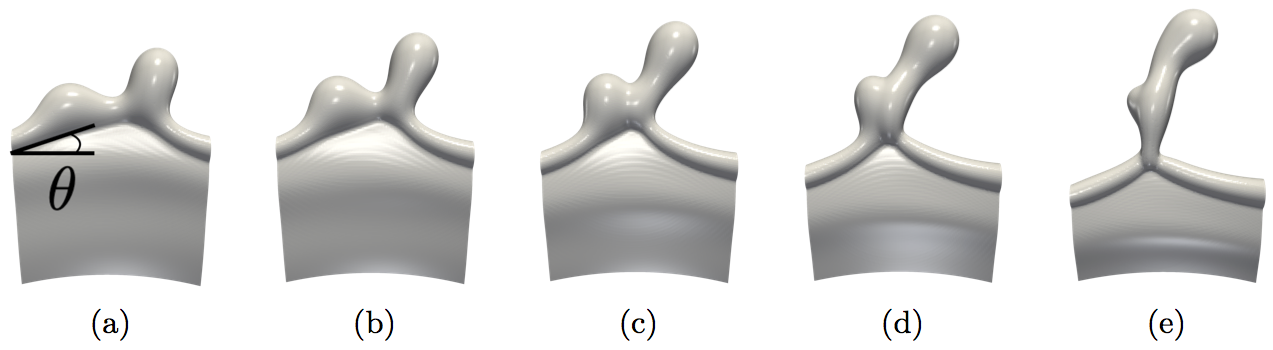} 
\end{center} 
\caption{\label{drift_velocity_figure}
Spatio-temporal evolution of two adjacent rim  protrusions which shift in the spanwise direction induced by local cusps, resulting in their collision; $\theta$ denotes the angle made by the left crest with the rim. The time difference between snapshots is $\Delta  t=1.0$.  The parameters remain unchanged from figure \ref{rim}.} 
\end{figure}

\begin{figure}
\begin{center} 
\includegraphics[width=1.0\linewidth]{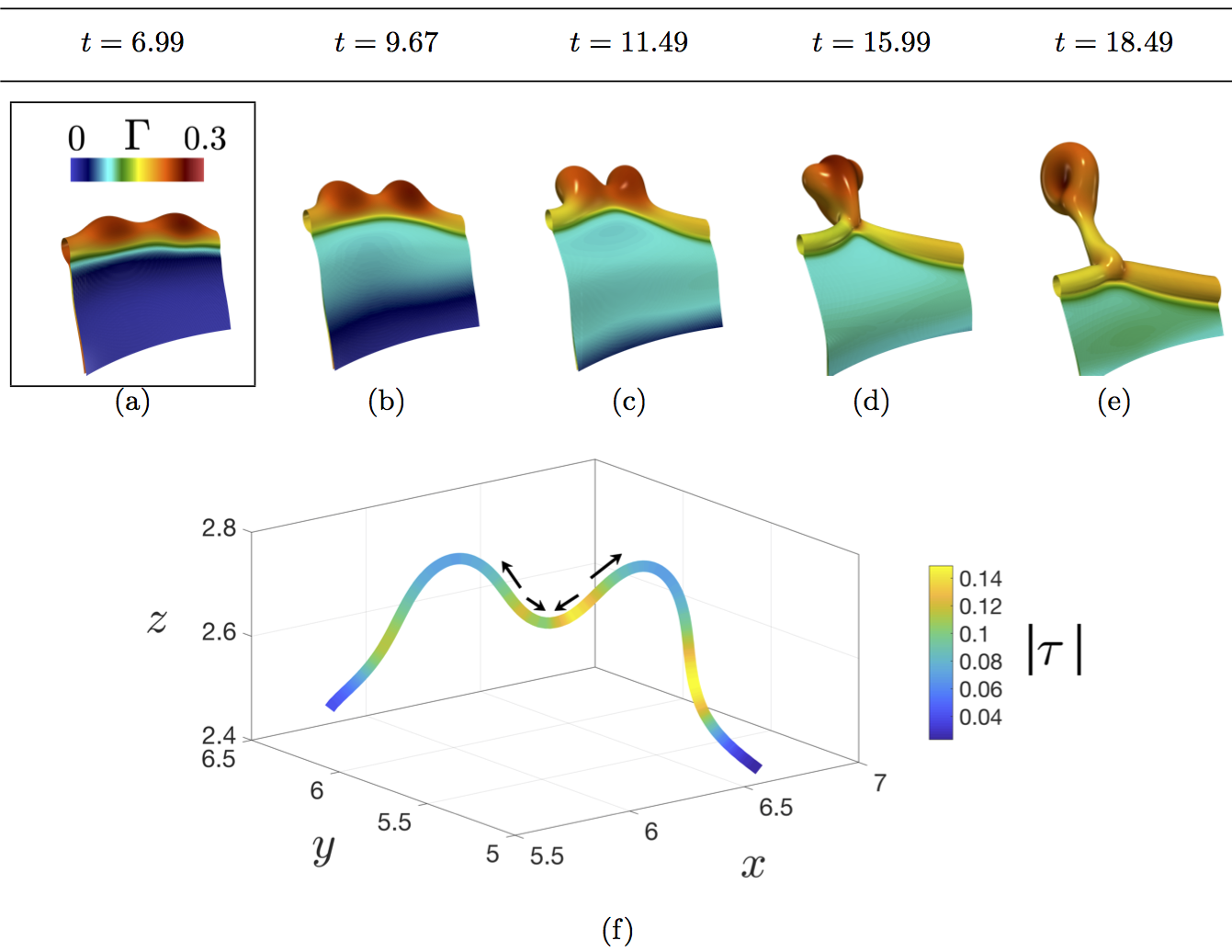} 
\end{center} 
\caption{\label{ligament_merge} Spatio-temporal evolution of two adjacent rim  protrusions shown panels $(a)-(e)$ which shift in the spanwise direction due to surfactant-induced Marangoni stresses, resulting in their  collision and merging. Panel $(f)$ shows a  three-dimensional reconstruction of the $|\tau|$  profile with respect to the arc-length, $s$, across a plane cutting the rim in (a) in half; the arrows represent the direction of $|\tau|$. Here, $\beta_s=0.5$, and all other  parameters remain unchanged from  figure \ref{temporal_evolution}.} 
\end{figure}

Next, we focus on drop formation and the splash phenomena: we have identified  three different modes of droplet shedding. The first mode corresponds to droplet detaching from the ligament via end-pinching.
\textcolor{black}{The mean lengths of the ligament $<l>$ which undergoes mode-1 of ejection are: $<l>=0.775$, $<l>=0.864$, $<l>=0.992$, and $<l>=1.141$, which correspond to the surfactant-free and surfactant-laden cases with $\beta_s=(0.1,0.5,0.7)$, respectively. This agrees with a surfactant-driven retardation of the capillary singularity.}
In the second mode we observe the initial ejection of filaments that are rapidly recombined into a single filament, see figure \ref{ligament_merge}(a) to (e). The third mode sees the formation of satellite (smaller) droplets during the stretching of the filament neck. These satellite droplets are not observed for large values of the elasticity parameter, in agreement with \citet{Kovalchuk_jcis_2018}. For the surfactant-free case, figure \ref{drift_velocity_figure}, ligaments merge  due to the presence of local cusps arising from a non-uniform distribution of mass per unit length  (in agreement with  \citet{gordillo_lhuissier_villermaux_2014} and \citet{wang_bourouiba_2018}), or by the rim parameter not being a multiple of the RP instability. In the vicinity of a cusp, the rim is not perpendicular to the flow entering the rim; instead, it takes an angle, $\theta$, as shown in figure \ref{drift_velocity_figure}a. 
This, oblique flow induces a local drift velocity along the rim. In a reference frame along the rim, the drift velocity can be calculated from the projection of the incoming velocity $u_{\rm {in}}$ in the longitudinal direction of the rim as $u_{\rm{drift}}=u_{\rm{in}}(t) \sin (\theta)$ (see \citet{wang_bourouiba_2018} for more details). Under the conditions of figure \ref{drift_velocity_figure}, we obtain $u_{\rm{drift}} \sim 0.106$ m/s and $\theta \sim 20^{\circ} \pm 3^{\circ}$. In contrast, from simulations, the velocity field yields $u_{\rm{in}} \sim 0.27$ m/s which leads to $u_{\rm{drift}} \sim 0.083$ m/s, and a good agreement with the direct measurement. The collision of the drifting protrusions results in the formation of a corrugated ligament (see figure \ref{drift_velocity_figure}e), which succumbs to end-pinching resulting in drop formation (not shown). We confirm that we have made the same calculation for the other two flow-induced cuspids resulting in similar results.

The presence of surfactant enhances the ligament-merging mode due to surfactant-driven Marangoni stresses that induce a local motion along the rim, and a radial shift  of the ligament position along the rim. This is highlighted in Figure \ref{ligament_merge}, which presents the temporal evolution of two  protrusions that  eventually  become ligaments, and merge prior to pinchoff.

Figure \ref{ligament_merge}a \textcolor{black}{demonstrates} that the highest $\Gamma$ values are found at the protrusions as the fluid motion along the rim to the protrusions  acts to the accumulation of surfactant locally. They are characterised by regions of a radially converging dynamics, and subsequently, increased (reduced) $\Gamma$ ($\sigma$) locally. These results demonstrate that the effect of Marangoni stresses from adjacent protrusions is to bridge their gap promoting the collision and merging of the ligaments. Figure \ref{ligament_merge}f shows a three-dimensional reconstruction of the $|\tau|$  profile with respect to the arc-length, $s$, across a plane cutting the rim seen in \ref{ligament_merge}a in half; the arrows represent the direction of $|\tau|$ evidencing the enhancement of the filament merging. To the best of our knowledge, this surfactant-induced mechanism of droplet shedding has not been reported previously. Figure \ref{ligament_merge} also shows that surfactant is highly convected from the rim to the film. Consequently, the  crown is thinner and survives longer in the presence of surfactants, in agreement with the experimental observations reported by  \citet{che_matar}.

%%%%%%%%%%%
%Distribution of droplet sizes
%%%%%%%%%%%

\begin{figure}
\begin{center}
\includegraphics[width=1.0\linewidth]{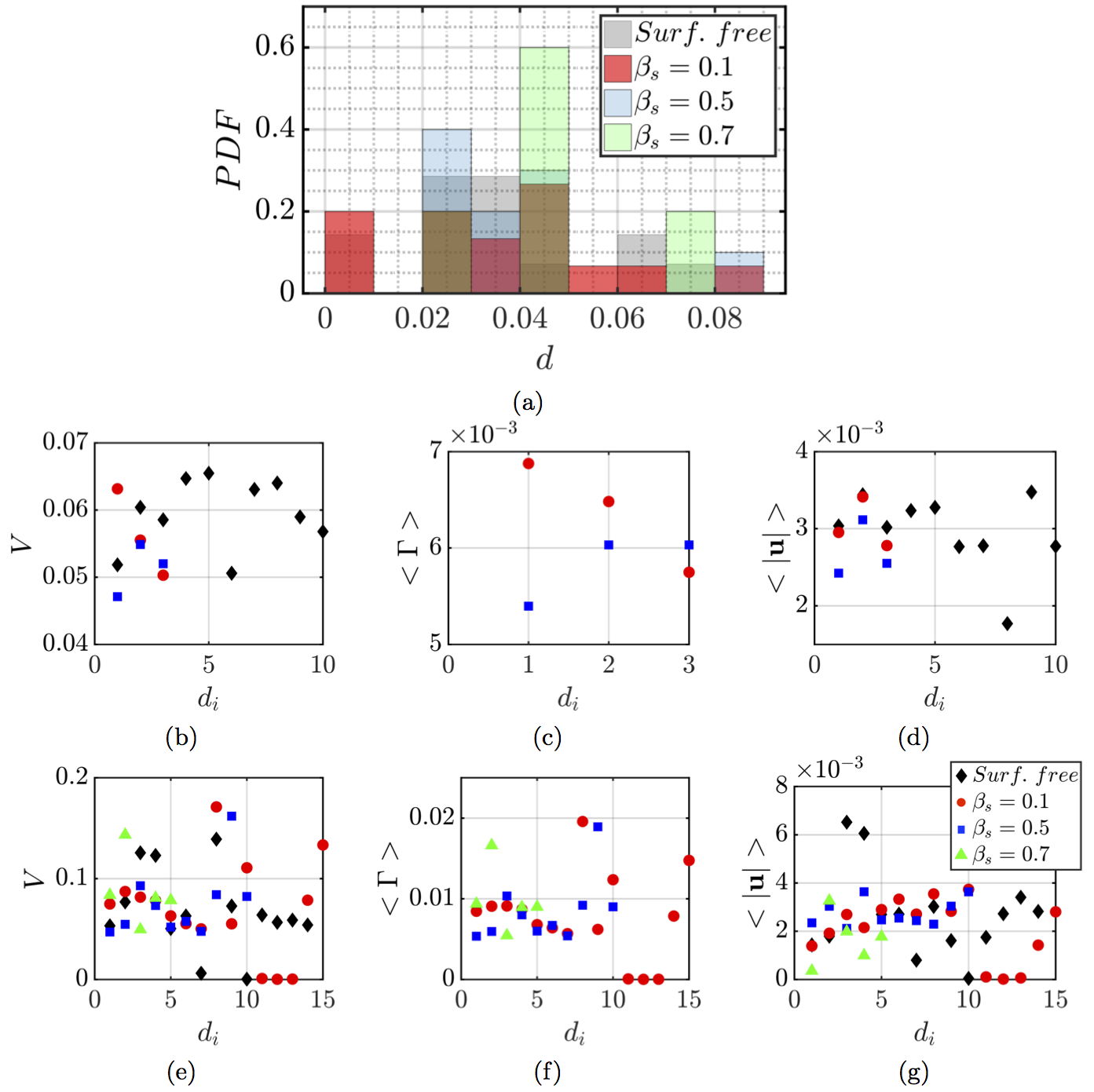} 
\end{center} 
\caption{\label{droplets} Metrics of splashing as a function of the elasticity parameter $\beta_s$. Panel (a) shows the probability density function ($PDF$) of the droplet size; droplet sizes are made dimensionless by the initial droplet diameter \textcolor{black}{$D_0$}. Panels $(b)-(d)$  and $(e)-(g)$ show the volume,  average surfactant concentration, and  average velocity for every droplet at $t=15$, and $t=20$, respectively ($d_i$ stands for the enumeration of droplets). All other  parameters remain unchanged from figure \ref{temporal_evolution}.} 
\end{figure}

The total number of ejected droplets for our surfactant-free case is found to be approximately $30$ up to times $t < 22.5$. In contrast, the number of secondary droplets (at $We,Re=800,1000$) as reported by \citet{agbaglah_deegan_2014}, correspond to $30-32$ so our numerical predictions are in good agreement. This also verifies that a RP instability is the primary mechanism for wavelength selection.

Next, we turn our attention to the size distribution of the droplets ejected from impact.  Figure \ref{droplets}a shows the probability density function ($PDF$), at $t=20$, scaled by the initial droplet diameter. The sampling is conducted after the impact reaches $t=20$ and over three snapshots separated by  time intervals of $\Delta t=0.2$. To get a smoother PDF, more snapshots should be included, but the dynamics are too rapid and due to the extreme computational cost of the current numerical simulation, this task was not undertaken. It is important to note that as $\beta_s$ increases, a higher droplet size is favoured as seen in Figures \ref{droplets}b-d and \ref{droplets}e-g where the droplet volume, average surfactant concentration, and average velocity for each droplet at $t=15$ and $t=20$, respectively, are shown. The  average velocity and surfactant concentration are calculated as $<|\bf{u}|>=\int_{\Omega }|\bf{u}| d \Omega  /  \Omega $ and $<\Gamma>=\int_{\Omega }\Gamma d\Omega /  \Omega $, respectively, where $\Omega$ represents the surface area of the droplet. We observe that fewer droplets are produced with increasing  $\beta_s$ is, fewer droplets are produced due to the rigidification brought by surfactant-driven Marangoni stresses (see figure \ref{droplets}b,e). For the high values of $\beta_s$, we observe some droplets with large volume corresponding to the merging of ligaments. We also observe that small droplets are produced for the surfactant-free and the $\beta_s = 0.1$ surfactant-laden cases, (see figure \ref{droplets}e). These small droplets correspond to satellite droplets, which are characterised by  $V, <|{\bf u}|>$ and $<\Gamma> \xrightarrow{} 0$, in good agreement with \citet{Kovalchuk} and \citet{che_matar}. We conclude that the addition of surfactants leads to a larger droplet size distribution for two reasons: the retardation of the dynamics, and the suppression of end-pinching through the reopening of the neck driven by a surfactant-induced flow. 

Figures \ref{droplets}(c,f) indicate that the average $<\Gamma>$ of the droplets is lower than the initial surfactant concentration of the pool (viz. $\Gamma=0.5$). This agrees well with the experimental studies of  \citet{Blanchard_1972} and \citet{constanteamores2020bb} on bursting bubbles. Figure \ref{droplets}c highlights that $<\Gamma>$ increases with decreasing $\beta_s$ as the surfactant has been highly convected towards the rim. Lastly, figures \ref{droplets}(d,g), which show $<|{\bf u}|>$ at $t=15$ and $t=20$, respectively, demonstrate the  effect of rigidification brought about by surfactant-induced Marangoni flow which corresponds to  an overall reduction in  $<|{\bf u}|>$ for the surfactant-laden droplets.

\begin{figure}
\begin{center} 
\includegraphics[width=1.0\linewidth]{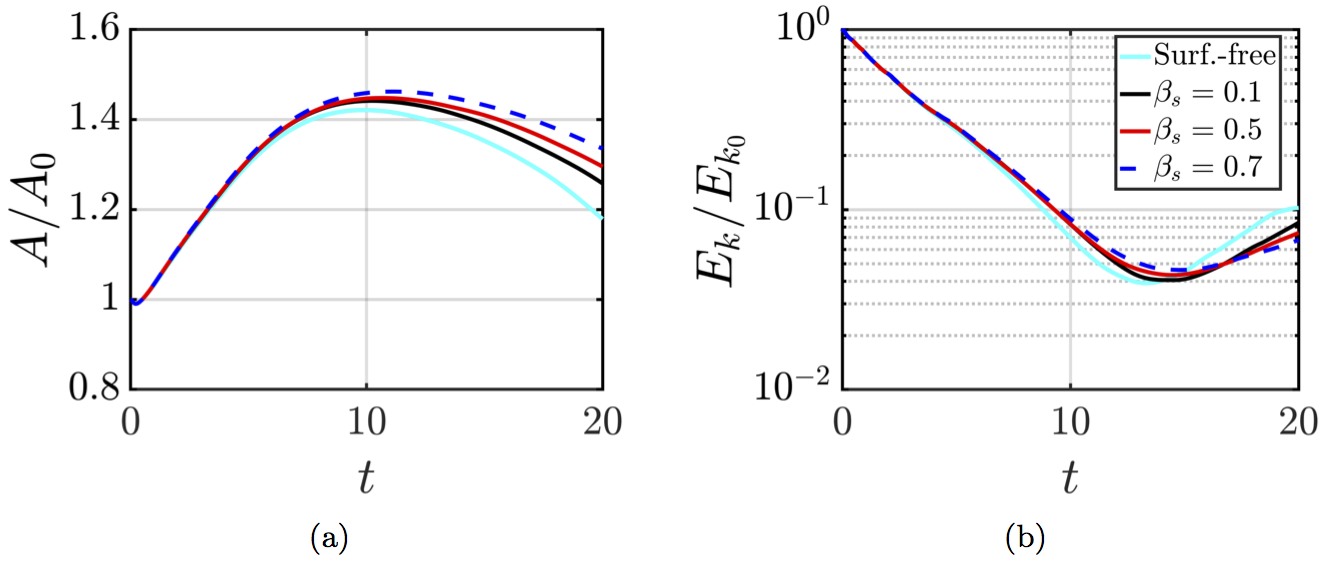} 
\end{center} 
\caption{Temporal evolution of the total interfacial area, (a),  and the kinetic energy, (b), normalised by their initial values, for  the surfactant-free and surfactant-laden cases. All  parameters remain unchanged from figure \ref{temporal_evolution}. \label{metrics}} 
\end{figure}

Lastly, we plot in figure \ref{metrics} the effect of surfactants on the interfacial area, and kinetic energy  defined as $E_k=\rho\int_V {\bf u}^2/2 dV$,  normalised by their initial values. Inspection of the surface area plot in figure \ref{metrics}a reveals a linear growth rate in interfacial area for all cases at short times; at longer times, a monotonic increase of surface area is observed with increasing $\beta_s$ due to surfactant-induced effects discussed above. In \ref{metrics}b, we see that the presence of surfactants by increasing $\beta_s$ acts to decrease the overall value of $E_k$ (e.g., rigidification brought about surfactants).

\section{Concluding remarks}\label{sec:Con}

We have presented, for the first time, a detailed analysis of the effect of insoluble surfactants on the dynamics of droplet impact on thin-film liquid layers using high-resolution three-dimensional simulations. This article focuses on the crown-splash regime, in which the transverse destabilisation of the rim leads to the formation of ligaments and droplets. Thus, we have focused on conditions of high Reynolds and Weber numbers. The nature of the hybrid front-tracking level-set numerical simulations enables the coupling, and analysis, of inertia, capillarity, interfacial diffusion, and Marangoni stresses owing to the presence of surfactant-induced surface tension gradients. According to our results,  the addition of surfactants does not significantly affect the selection of the wavelength of the transversal rim instability; as the predicted wavelength is consistent with the most unstable Rayleigh-Plateau instability. The instability give rise to the formation of ligaments which eventually end-pinch into droplets. In the presence of surfactants, we observe a delay in end-pinching, driven by surfactant-induced Marangoni stresses, resulting in longer ligaments. We identify three modes of droplet ejection, and demonstrate that surfactant-induced Marangoni stresses result in the bridging of the spanwise surfactant concentration gradients between adjacent ligaments in the rim, eventually leading to their merging. The presence of surfactant-induced Marangoni stresses also leads to interfacial rigidification, which is observed through a reduction of the surface velocity and the kinetic energy, and a larger interface area.

This research is of importance to many applications as droplet impact onto finite-depth liquid layers is found in a wide range of industrial and daily-life applications, from raindrops impinging onto puddles to screen and inkjet printing. This research focuses on the crown splashing regime but we acknowledge that the presence of surfactants will play a crucial role in other $We-Re$ regions, e.g. at the microdoplet-splash regime. Although we have only focused on insoluble surfactants,  the presence of surfactant-solubility will lead to additional richness and complexity, and we anticipate that the addition of soluble surfactants may result in detrimental effects, and they will be the subject of future work. Additionally, surfactant-induced rheological effects could be important and of great relevance to industry and most industrial fluids are formulated with dispersants (surfactants). 

Finally, it is well-known that the thickness of the film layer affects the interfacial dynamics. Additionally, most of literature has focused on  the limiting case of  $t\ll D/U$ in which axisymmetric simulations suffice. The main novelty of this work is the study of the role of the surfactant at longer times after rim formation. At these times, surfactants play a major role in the late stages of droplet detachment, and, to the best of our knowledge, this manuscript is the first computational study on this phenomenon. Consequently, future studies could focus on the surfactant-induced effects of droplet impacting onto various liquid film thicknesses, and the ejecta sheet.\\

%\subsection*{Acknowledgements}
Declaration of Interests. The authors report no conflict of interest. \\

CRC-A and LK thank with gratitude Dr A. Batchvarov for the fruitful discussions. AAC-P acknowledge the support from the Royal Society through a University Research Fellowship (URF/R/180016), an Enhancement Grant (RGF/EA/181002) and two NSF/CBET-EPSRC grants (Grant Nos. EP/S029966/1 and EP/W016036/1). O.K.M. acknowledges funding from PETRONAS and the Royal Academy of Engineering for a Research Chair in Multiphase Fluid Dynamics, and the EPSRC MEMPHIS (EP/K003976/1) and PREMIERE (EP/T000414/1) Programme Grants.  We acknowledge HPC facilities provided by the Research Computing Service (RCS) of Imperial College London for the computing time. D.J. and J.C. acknowledge support through computing time at the Institut du Developpement et des Ressources en Informatique Scientifique (IDRIS) of the Centre National de la Recherche Scientifique (CNRS), coordinated by GENCI (Grand Equipement National de Calcul Intensif) Grant 2022 A0122B06721. 
This research was funded in whole or in part by the UK EPSRC grant numbers specified above. For the purpose of Open Access, the author has applied a CC BY public copyright licence to any Author Accepted Manuscript (AAM) version arising from this submission.

\section*{Appendix A}

\begin{figure}
\begin{center} 
\includegraphics[width=1.0\linewidth]{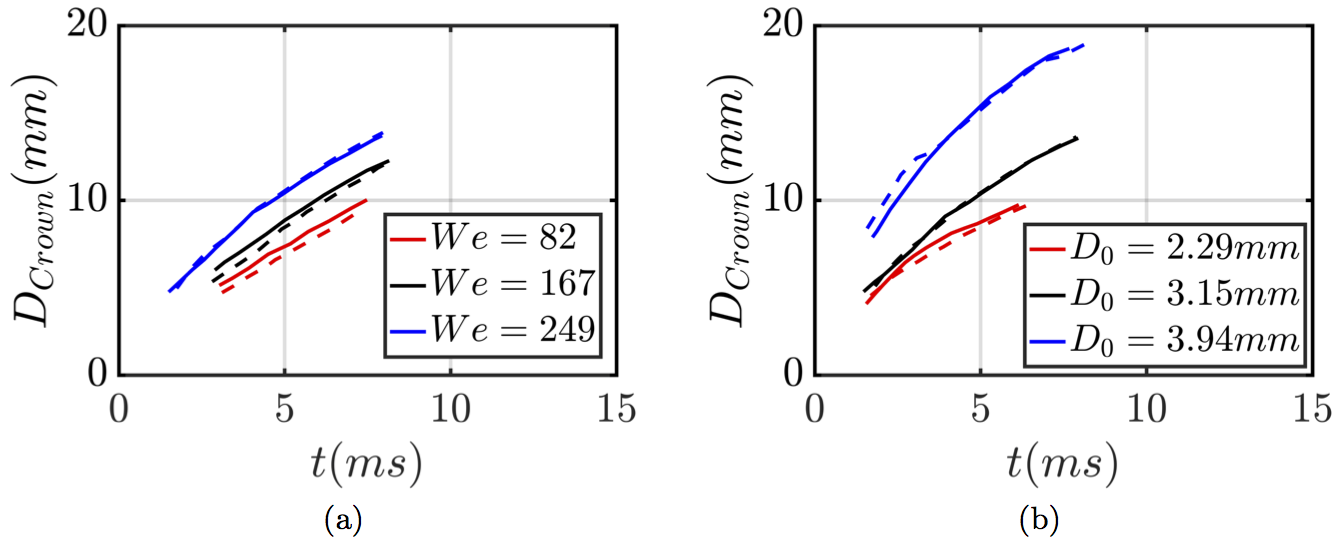} 
\end{center} 
\caption{\label{extra_validation} Additional validation of the numerical framework  (solid lines) against the experimental data  (dashed lines) of \citet{che_matar}. (a) sees the effect of varying the Weber number and (b) shows the effect of varying droplet size on the crown diameter.}
\end{figure}

Figure \ref{extra_validation} a,b show additional numerical results against the experimental data of Che \& Matar (2017) for the crown rim evolution. Here, the  effect of varying the Weber number, e.g. $We=(82,157,249)$, at a constant film depth $ h=0.22$, $Re=(4312,6153,7514)$ and $Fr=(7.96, 10.80,13.65)$, is seen for various droplet sizes. Due to the difference in droplet size, the dimensionless groups vary between each case: a small droplet with a diameter $D=0.00229$ m is described by $We=183$, $Re=5411$, $Fr=16.01$, and $ h=0.13$; a medium droplet with diameter $D=0.00315$ m is characterised by $We=249$, $Re=6311$, $Fr=13.65$, and $ h=0.095$; a large droplet with $D=0.00394$ m is represented by $We=315$, $Re=9341$, $Fr=12.20$, and $ h=0.076$. We observe good agreement between our numerical predictions and the reported literature; we conclude our numerical approach faithfully captures the rich dynamics of droplet impact onto a thin film layer of the same liquid.

\section*{Appendix B}\label{sec:App}

The code has been benchmarked against: i) simulations of \citet{notz_basaran_2004} on the recoiling of surfactant-free liquid threads, ii) the Rayleigh-Plateau instability model for liquid jets of \citet{Plateau_1873}, iii) the scaling laws of pinching off threads by \citet{Eggers_prl_1993} and \citet{Lister_Stone_pf_1998}, and iv) the work of \citet{Blanchette2006} on droplet impact at vanishing velocity (droplet coalescence on a flat surface). These validations are found elsewhere, i.e. \citet{constante_jets,Constante-Amores_prf_2020,constante_2023}.

\begin{figure}
\begin{center} 
\includegraphics[width=1.0\linewidth]{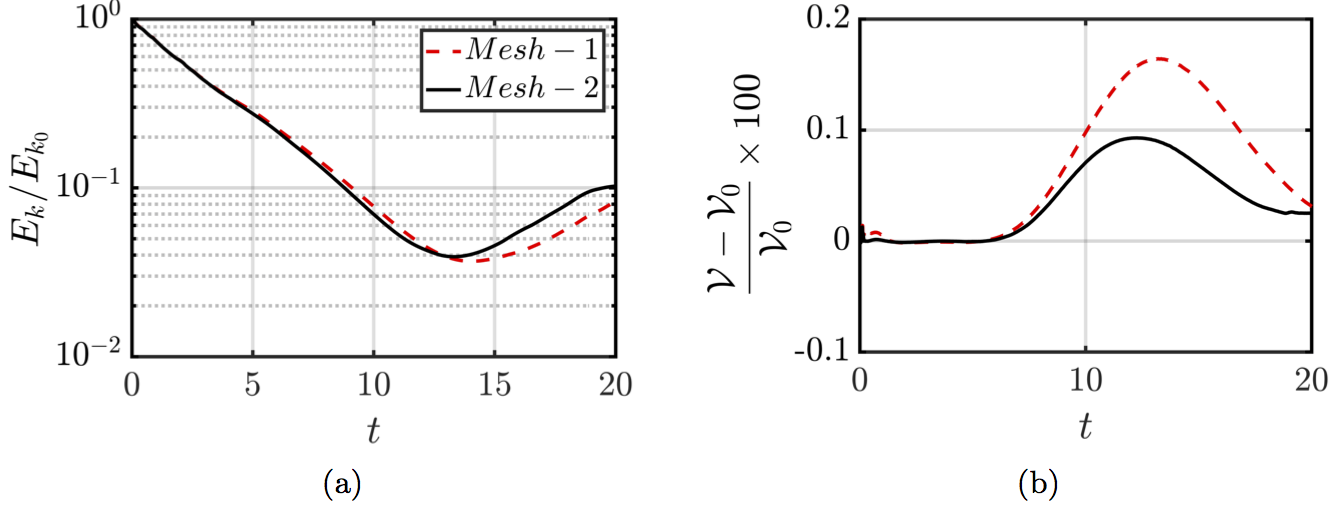}
\end{center} 
\caption{Mesh study for the surfactant-free case when $Re=1000$ and $We=800$. The panels highlight the temporal evolution of  kinetic energy $E_k$, and the relative variation of the liquid volume for two different meshes \label{mesh}} 
\end{figure}

Figure \ref{mesh} displays the temporal evolution of the kinetic energy, $E_k$, and the conservation of liquid volume for the surfactant-free case ($Re=1000$ and $We=800$) for different mesh resolutions. `Mesh-1' and  `Mesh-2' refer to different levels of refinement characterised by $384^2\times192$, and $768^2\times 384$, respectively. As demonstrated in figure \ref{mesh}, both levels of refinement can  reproduce the dynamics with a conservation of fluid volume around $0.1\%$. Based on these results, we conclude that  `Mesh-2' is sufficiently refined to ensure mesh-independent results while providing a good compromise with the computational cost. All the results presented in this paper correspond to a `Mesh-2' mesh.

\end{document}